\documentclass[10pt,aps,twocolumn]{revtex4}

\usepackage{amsfonts}
\usepackage{graphicx}
\usepackage{amsmath}
\usepackage{amsthm}
\usepackage[colorlinks=true, citecolor=blue, urlcolor=blue ]{hyperref}
\usepackage{graphicx}

\begin{document}
\title{Complementary correlations and entanglement distribution}

\author{Prasenjit Deb}
\email{devprasen@gmail.com}
\affiliation{Department of Physics and Center for Astroparticle Physics and Space Science, Bose Institute, Bidhan Nagar
Kolkata - 700091, India.}

\begin{abstract}
Complementary correlations can reveal the genuine quantum correlations present in a composite quantum system. Here we investigate
the relation between complementary correlations and other aspects of genuine quantum correlations. We show that for a certain class 
of states quantum correlations revealed through complementary correlations become equal to entanglement and discord. We also provide a necessary and sufficient condition
for entanglement distribution with separable Bell diagonal states in terms of complementary correlations. 
\end{abstract}
\maketitle

\section{Introduction}
Quantum mechanical systems consisting of more than one subsystem can have both quantum and classical correlations
or only classical correlations between themselves\cite{Vedral}.There are different aspects of quantum correlations, 
such as entanglement\cite{Schrodinger,Horodecki}, discord\cite{Zurek},
measurement-induced disturbances\cite{luo3} etc.The presence of quantum correlations make the quantum correlated states 
useful for quantum information processing tasks such as \emph{teleportation}\cite{Tele},\emph{dense-coding}\cite{Dense}, 
\emph{remote state preparation}\cite{RSP},
\emph{quantum cryptography}\cite{chuang} etc.So it is very important to characterize and quantify such non-classical correlations. 
Though entanglement is the best studied form of quantum correlations, in the last few years discord has also received 
a lot of attention. 
\paragraph*{}
Apart from entanglement and discord,
the quantum correlations can also be revealed by measuring the correlations between measurement outcomes of
complementary observables present on both sides of a bipartite state\cite{Lorenzo}, say $\rho^{AB}$.  
Though complementary correlation measures can be linked to mutual information, Pearson correlation coefficient and the sum of 
conditional probabilities\cite{Lorenzo}, we are mainly interested on mutual information based measure. 
In this article we show that for certain class of states the quantum correlations measured through 
complementary correlation is exactly equal to entanglement and discord.

\paragraph*{}
On the other hand, entanglement distribution with seperable states is a quantum phenomena which depicts
the bizzare feature of quantum correlations.First introduced by Cubitt$\it{et.al.}$\cite{cubitt}, 
this phenomena has got a lot of attention
in the last few years.The central idea of entanglement distribution scheme in Ref.\cite{cubitt} 
is that Alice can create entanglement between herself
and Bob by sending a qubit in a seperable state.At the beginning of the protocol  
Alice and Bob take a state $\rho_{ABC}$ which 
is shared between them as $\rho_{AC\lvert B}$. Alice then applies a controlled NOT (CNOT) on her qubits $A$ and $C$
(where $A$ is the control qubit) and sends qubit $C$ to Bob.
Upon receiving qubit $C$ from Alice, Bob performs CNOT on qubits $B$ and $C$ taking $B$ as the control qubit. At the end
of the protocol entanglement is generated in the bipartition $\rho_{A\lvert BC}$. During the whole process qubit $C$ remains 
separable from the labs of Alice and Bob, that is bipartitioning $\rho_{C\lvert AB}$ is seperable.

\paragraph*{}
Though Alice and Bob can create entanglement from a seperable state shared by them, the amount 
of entanglement gain is always upper bounded which simply means that arbitrary amount of entanglement can not be generated. 
More precisely to say, the non-classical correlations of the carrier (qubit $C$) with parties( qubits $A$ and $B$) 
bound the amount of distributed entanglement\cite{dagmar,chuan}. 
\paragraph*{}
The protocols for entanglement distribution described in Ref\cite{dagmar,chuan} have one common point. All of those protocols assume 
that initially there is some quantum correlations between ancilla qubit $C$ and the laboratories. But here we are interested 
with such a protocol where ancilla qubit is uncorrelated with rest of the system, as also mentioned in Ref\cite{kay}. 
Keeping such a protocol in mind if we think EDSS as a quantum task in which entanglement is activated from other form of 
quantum correlations, then seperable states are the useful resources in the task and ancilla qubit helps in the distribution 
of entanglement. At this point one might be tempted to ask the question- 
\emph{What are the necessary conditions a seperable state must satisfy to be useful  for entanglement distribution}? 
The answer to this question has been provided in \cite{dagmar1}, where it was shown that entanglement distribution with seperable states is possible if the rank of the states are at least 3. 
 We aim here to find the answer of the above question through a different approach. More specifically, imposing conditions on quantum correlations present 
in the seperable state. We consider complementary correlations as the measure of quantum correlations and derive necessary 
and sufficient condition for EDSS.

\section{Complementary Correlations}
Two quantum mechanical observables are called complementary if knowledge of the measured value any one observable implies
maximal uncertainty of the measured value of another. To explain more precisely, let $\mathcal{A}$ and $\mathcal{B}$ are two
non-degenerate observables. The spectral representation of these observables are respectively $\mathcal{A}= \Sigma_i~ f(a_i)\lvert a_i\rangle\langle a_i\lvert$
and $\mathcal{B}= \Sigma_j~ g(b_j)\lvert b_j\rangle\langle b_j\lvert$, where$\lvert a_i\rangle$ and $\lvert b_j\rangle$ are
the eigenstates of $\mathcal{A}$ and $\mathcal{B}$ respectively. $f$ and $g$ are two arbitrary bijective functions. Now, the 
observables $\mathcal{A}$ and $\mathcal{B}$ are complementary if 
\begin{equation}
 \lvert\langle a_i\lvert b_j\rangle\lvert = \frac{1}{d},~~~~~ \forall i,j
\end{equation}
where, $d$ is the dimension of the Hilbert space of the quantum system.
The complementary observables with the above definition identify two mutually unbiased bases(MUBs)\cite{som}. For example, in 2-dimensional
Hilbert space the  Pauli spinors $\sigma_x$,$\sigma_y$ and $\sigma_z$ are the three complementary observables,
the eigenvectors of which form the mutually unbiased bases $\{\lvert0\rangle,\lvert1\rangle\}$,
$\{\frac{\lvert0\rangle+\lvert1\rangle}{\sqrt{2}}, \frac{\lvert0\rangle-\lvert1\rangle }{\sqrt{2}}\}$ and 
$\{\frac{\lvert0\rangle+i\lvert1\rangle}{\sqrt{2}}, \frac{\lvert0\rangle-i\lvert1\rangle }{\sqrt{2}}\}$

\paragraph*{}

Now consider Alice and Bob are two parties holding quantum systems $A$ and $B$ respectively, $\{\mathcal{A}_i\}$ and
$\{\mathcal{B}_i\}$ are respectively the set of complementary observables for Alice's and Bob's systems. 
The correlations between the measurement 
outcomes of complementary observables $\mathcal{A}_i$'s and $\mathcal{B}_i$'s can be expressed in terms mutual information as\cite{Lorenzo}
\begin{equation}
 I(\mathcal{A}_i:\mathcal{B}_i)\equiv H(\mathcal{A}_i) -H(\mathcal{A}_i\lvert \mathcal{B}_i)
\end{equation}
where, $H(\mathcal{A}_i)$ is the Shanon entropy of the probabilities of the measurement outcomes of observable $\mathcal{A}_i$
of Alice's system and $H(\mathcal{A}_i\lvert \mathcal{B}_i)$ is the conditional entropy, conditioning being done on Bob's
side. If $\mathcal{A}_1$,$\mathcal{A}_2$ and $\mathcal{B}_1$,$\mathcal{B}_2$ are complementary observables on Alice's and Bob's 
side, then the bipartite quantum state shared between Alice and Bob is maximally entangled if and only if
$I(\mathcal{A}_1:\mathcal{B}_1)$ +$ I(\mathcal{A}_2:\mathcal{B}_2)= 2 \log_2 d$\cite{Lorenzo},where $d$ is the dimension of the Hilbert space
of Alice's or Bob's system.
\paragraph*{}
We begin with a generic two-qubit state \cite{Horodecki1} which can be  represented as
\begin{equation}
 \tau=\frac{1}{4}(\mathbf{1}_2\otimes\mathbf{1}_2+\mathbf{a}\cdot\boldsymbol{\sigma}\otimes\mathbf{1}_2 +\mathbf{1}_2
 \otimes \mathbf{b}\cdot\boldsymbol{\sigma} +\sum _{m,n=1}^{3}c_{nm}\sigma_n\otimes\sigma_m )\label{2qubit}
\end{equation}
where, $\mathbf{1}_2$ represents the identity operator, $\mathbf{a}$ and $\mathbf{b}$ are the local Bloch vectors for each subsystem and $\{\sigma_n\}_{n=1}^{3}$ are the standard Pauli spinors $\sigma_x$,
$\sigma_y$ and $\sigma_z$. The coefficients $c_{nm} =\mbox{Tr}(\tau\sigma_n\otimes\sigma_m)$ form the elements of 
a $3\times3$ matrix $\mathcal{T}$.
From\cite{Horodecki1,luo2} it is easy to show that $\tau$ is locally unitary equivalent to
\begin{equation}
 \gamma = \frac{1}{4}(\mathbf{1}_2\otimes\mathbf{1}_2+ \mathbf{a}\cdot\boldsymbol{\sigma}\otimes\mathbf{1}_2+
 \mathbf{1}_2\otimes \mathbf{b}\cdot\boldsymbol{\sigma}+\sum _{n=1}^{3} c_{n}\sigma_n\otimes\sigma_n ).
\end{equation}
For our purpose  we will consider the states with maximally mixed marginals,that is, the states of the form
\begin{equation}
 \rho = \frac{1}{4}(\mathbf{1}_2\otimes\mathbf{1}_2+\sum _{n=1}^{3}c_{n}\sigma_n\otimes\sigma_n),\label{bell-diag}
\end{equation}
As both the subsystems are qubit, the eigen vectors of Pauli spinors form the mutually unbiased bases  on Bob's side, which are:
\begin{eqnarray}
 \{\lvert b^j_1\rangle\lvert j=1,2\} &=&\{\lvert0\rangle,\lvert1\rangle\}\nonumber\\
 \{\lvert b^j_2\rangle\lvert j=1,2\} &=&\{\frac{\lvert0\rangle+\lvert1\rangle}{\sqrt{2}}, \frac{\lvert0\rangle-\lvert1\rangle }{\sqrt{2}}\}\nonumber\\
 \{\lvert b^j_3\rangle\lvert j=1,2\} &=&\{\frac{\lvert0\rangle+i\lvert1\rangle}{\sqrt{2}}, \frac{\lvert0\rangle-i\lvert1\rangle }{\sqrt{2}}\}
\end{eqnarray}
Our primary goal is to measure quantum correlations by measuring the complementary correlations of the state $\rho$. Let
\begin{equation}
 \{\Pi_{\{\lvert b^j_i\rangle\lvert j=1,2\}} = \lvert b^j_i\rangle \langle b^j_i\lvert: i=1,2,3;j=1,2\}
\end{equation}
be the local projective measurements on Bob's side. Interestingly, each measurement is related to other through some 
unitary $V\in U(2)$, that is, 
\begin{equation}
 \Pi_{\{\lvert b^j_2\rangle\lvert j=1,2\}} = V\Pi_{\{\lvert b^j_1\rangle\}}V^{\dagger}
\end{equation}
and so on. Here $V$ has an explicit form as
\begin{equation}
 V= tI+i\vec{y}\vec{\sigma}
\end{equation}
where, $t\in\mathcal{R}$, $\vec{y}=(y_1,y_2,y_3)\in\mathcal{R}^3$. Now, using the same techniques as in Ref.\cite{luo2}, we find 
the complementary correlations as
\begin{eqnarray}
\chi\{\rho\lvert\{\Pi_{\{\lvert b^j_i\rangle\lvert j=1,2\}}\}\}&=&S(\Sigma_i~ p_i\rho_i^A)-\Sigma_i~p_iS(\rho_i^A)\nonumber\\
&=&\frac{1+\theta}{2}\log(1+\theta)\nonumber\\
&&+\frac{1-\theta}{2}\log(1-\theta)\label{maxholevo}
\end{eqnarray}
where, $\theta\equiv\theta(t,y_i,c_i)$. The maximum of the quantity in Eqn.({\ref{maxholevo}}) will give the value of classical 
correlation of the state. Let, for $\theta=\lvert c_1\lvert$ the quantity in Eqn.({\ref{maxholevo}}) becomes maximum. So the classical correlation
of the state wil be
\begin{equation}
 \mathcal{C}(\rho)= \frac{1+c_1}{2}\log(1+c_1)+\frac{1-c_1}{2}\log(1-c_1)\label{cc}
\end{equation}
A little algebra shows that $\theta=\lvert c_1\lvert$ actually refer to measurement on 
$\{\lvert b^j_2\rangle\lvert j=1,2\}=\{\frac{\lvert0\rangle+\lvert1\rangle}{\sqrt{2}}, \frac{\lvert0\rangle-\lvert1\rangle}{\sqrt{2}} \}$
basis. 
In other words, we can say that the correlations between measurement outcomes of the observable $\sigma_x$ on both sides
will yield the classical correlation of the state $\rho$. The correlations between measurements outcomes
of the observables $\sigma_z$  will be
\begin{equation}
 \mathcal{Q}_1=\frac{1+c_3}{2}\log(1+c_3)+\frac{1-c_3}{2}\log(1-c_3)\label{com corr}
\end{equation}
The discord $(\mathcal{D})$ of the state $\rho$ can be found from Ref.\cite{luo2} and comparison of that with $\mathcal{Q}_1$
clearly reveals the relation
\begin{equation}
 \mathcal{Q}_1\leq\mathcal{D}
\end{equation}

If the total correlations or mutual information($\mathcal{I}$) of the state $\rho$ is calculated, then it is clear that
\begin{equation}
 \mathcal{Q}_1 + \mathcal{C} < \mathcal{I}
\end{equation}

\paragraph*{}
Now we consider the states for which $c_1=1$, $c_2=-c_3$ and $\lvert c_3\lvert\leq 1$. The states with such parametrization are of the form
\begin{equation}
 \rho= \frac{1+c_3}{2}\lvert\Psi^{+}\rangle\langle\Psi^{+}\lvert+\frac{1-c_3}{2}\lvert\Phi^{+}\rangle\langle\Phi^{+}\lvert \label{st}
\end{equation}
where, $\lvert\Phi^{+}\rangle= \frac{1}{\sqrt{2}}(\lvert 00\rangle+\lvert 11\rangle)$ and
$\lvert\Psi^{+}\rangle= \frac{1}{\sqrt{2}}(\lvert 01\rangle+\lvert 10\rangle)$
The discord and relative entropy of entanglement of such states are found to be
\begin{equation}
 \mathcal{D}= \mathcal{E}_r= \mathcal{Q}_1
\end{equation}
and the classical correlation is exactly the same as that in Eqn.({\ref{cc}}).
Hence, for the states described in  Eqn.({\ref{st}}) quantum correlations revealed through complementary correlation
is equal to other existing measure of non-classical correlations such as discord and relative entropy of entanglement,
and for such states
\begin{equation}
 \mathcal{Q}_1 + \mathcal{C} = \mathcal{I}
\end{equation}

\paragraph*{}
If one considers the Werner state\cite{Werner}, then it will not be hard to prove that $\mathcal{Q}_1\leq\mathcal{D}$. 
\paragraph*{}
Again consider a two-qubit state $\rho^{AB}$. If any of the complementary correlations $I(\sigma_x^A:\sigma_x^B)$,$ I(\sigma_y^A:\sigma_y^B)$
and $I(\sigma_z^A:\sigma_z^B)$ of the state is zero, then what can be infered about discord and entanglement of $\rho^{AB}$?
For any generic two qubit state it is hard to infer about discord and entanglement from complementary correlations but for Bell-diagonal states complementary correlations serve the purpose.
\subsection{A necessary condition for non-zero entanglement of Bell-diagonal states}
\emph{Bell-diagonal states will have non-zero entanglement if all the complementary correlations are non-zero.}\\\\\
Proof:-
Bell diagonal states are represented as
$\rho_{AB} = \frac{1}{4}(\mathbf{1}_2\otimes\mathbf{1}_2+\sum_ {n=1}^{3}c_{n}\sigma_n\otimes\sigma_n)$.  Let $\lambda(\rho_{AB})$ denote the spectrum of state $\rho_{AB}$.The partial transpose of $\rho_{AB}$ on qubit $A$ be  $\rho^A_{AB}$.
Now, let us consider that
\begin{equation}
 \mathcal{I}(\sigma_{(y)A}:\sigma_{(y)B})= 0
\end{equation}
which implies
\begin{equation}
\lvert c_2 \lvert=0
\end{equation}
Under such a condition $\lambda(\rho^A_{AB})=\lambda(\rho_{AB})\geq 0$ and PPT condition\cite{ppt} confirms that the state will be seperable and hence has zero-entanglement. Similarly, considering  any other complementary correlation to be equal to zero it can be proved that the state $\rho_{AB}$ will have zero-entanglement.
\paragraph*{}
The above mentioned condition is necessary but not sufficient.

\subsection{Complementary correlations and discord}
Again consider two Bell diagonal states of the form in Eqn.(\ref{bell-diag}) with $c_1=0.5,c_2=0.25,c_3=0.25$ and $c_1=0.5,c_2=0,c_3=0.25$ respectively. 
The discord of those states when calculated shows the relation $\mathcal{D}_{c_2=0.25}>\mathcal{D}_{c_2=0}$. Therefore, it 
is clear that Bell diagonal states for which all the complementary correlations are non-zero will have more discord than that of those
for which $I(\sigma_y^A:\sigma_y^B)= 0$ . 

\paragraph*{}
For classically correlated state of the form
\begin{equation}
 \rho_{cc}=\frac{1}{2}(\lvert 00\rangle\langle00\lvert+\lvert 11\rangle\langle11\lvert)
\end{equation}
 $I(\sigma_x^A:\sigma_x^B)= 0$, $I(\sigma_y^A:\sigma_y^B)= 0$ and  $I(\sigma_z^A:\sigma_z^B)= \mathcal{C}(\rho_{cc})$
 \paragraph*{}
From the above examples it is well understood that complementary correlations are very useful in revealing 
the genuine quantum correlations.

\section{Necessary and sufficient condition for entanglement distribution with seperable Bell-diagonal states}

Separable Bell diagonal states will be  useful resource for entanglement distribution \emph{iff} correlations 
exist between measurement outcomes of all the complementary observables present on both side. The proof of the statement 
is as follows:
\paragraph*{}
Let $\rho_{AB}$ is a Bell-diagonal state as represented in Eqn.(\ref{bell-diag}) and  $\lambda(\rho_{ABC})$ denote the spectrum of state $\rho_{ABC}$,
where, $\rho_{ABC}= U_{AC}\rho_{AB}\otimes \rho_C U_{AC}^ \dagger$ and $U_{AC}$ is the unitary applied by Alice 
on qubits $A$ and $C$.
The partial transpose of $\rho_{ABC}$ on qubit $A$ be  $\rho^A_{ABC}$. Entanglement distribution is said 
to be successful if in the bipartition $A\lvert BC$, $\lambda(\rho^A_{ABC}) < 0$\cite{boundentanglement}. 
\paragraph*{}
Now let 
$\mathcal{I}(\sigma_{(y)A}:\sigma_{(y)B})=0$, which implies $\frac{(1+c_2)}{2}\log_2(1+c_2)+\frac{(1-c_2)}{2}\log_2(1-c_2)=0$ and hence $ \lvert c_2\lvert  =0$. Under such a condition $\lambda(\rho^A_{ABC}) <0$ does not hold \cite{kay}, i.e., no distillable entanglement is present between Alice and Bob.
\paragraph*{}
Similarly if we consider $\mathcal{I}(\sigma_{(x)A}:\sigma_{(x)B})=0$ or $\mathcal{I}(\sigma_{(z)A}:\sigma_{(z)B})=0$ ( equivalently $\lvert c_1 \lvert =0$ or $\lvert c_3\lvert=0$) then $\lambda(\rho^A_{ABC})= \lambda(\rho_{ABC})$. Now, as $\rho_{ABC}$ is a well defined desity matrix representing a 3-qubit state, $\lambda(\rho_{ABC}) \geq 0$. Hence, there will be no distillable entanglement in the bipartition $A\lvert BC$.

\section{Conclusion}\label{sec4}
We have shown that for a certain class of states the genuine quantum correlations can be measured through complementary 
correlations. For such states the non-classical correlations measured through complementary correlations are exactly equal to
the entanglement and discord.
Considering Bell-diagonal states we have emphasized that simultaneous existence
of correlations in all the mutually unbiased bases(equivalently $\mathcal{I}(\sigma_{(x)A}:\sigma_{(x)B})\neq0$  
$\mathcal{I}(\sigma_{(y)A}:\sigma_{(y)B})\neq0$, $\mathcal{I}(\sigma_{(z)A}:\sigma_{(z)B})\neq0$)
is necessary for non-zero entanglement.
Moreover, for such states if correlations in any complementary base vanish then the quantum discord will decrease. 
Though we have considered a very special class of states and tried to investigate how complementary correlations can provide
insights about discord and entanglement, we aim to generalize the results for any bipartite 2-qubit states. 
\paragraph*{}
We have also provided a necessary and sufficient condition for entanglement distribution with seperable Bell-diagonal states.
A seperable Bell-diagonal state will be a useful resource for EDSS \emph{iff} all the complementary correlations are non-zero.
The condition for EDSS provided by us seems to be equivalent to that mentioned in Ref\cite{kay} but our approach is entirely
different. Infact, if the coefficients $c_is$  are ordered like $\lvert c_1\lvert \geq \lvert c_3\lvert \geq \lvert c_2\lvert$
and it is assumed that $\mathcal{I}(\sigma_{(x)A}:\sigma_{(x)B})=0$, then $\lvert c_2\lvert=0$ automatically. So, the 
necessary condition  $\lvert c_2\lvert \neq 0$ for EDSS with Bell-diagonal states\cite{kay} follows from the condition shown in this paper. Entanglement
distribution with seperable states depends on the quantum correlations present in such states and we hope that our results
will shed light on this fact.

\section{Acknowledgements}
This work is funded through INSPIRE-Fellowship Scheme by Department of Science and Technology, Govt. of India. The author acknowledges Professor Alastair Kay for providing useful insights.

\end{document}